# Universidad San Francisco de Quito
# La luz que perciben las abejas

André Morán

30 de Abril del 2020

**Resumen**

En este proyecto se profundiza en los conceptos de las ondas electromagnéticas de la luz y se introducen los conceptos de longitud de onda y polarización, se relacionan estos conceptos con la capacidad que tienen las abejas para percibir de luz ultravioleta y luz polarizada. Se estudia esta cadena causal de fenómenos físicos mediante una descripción detallada de las fuentes de luz y las formas en que se puede polarizar la misma en el entorno en el que viven las abejas. Además, se estudian los mecanismos anatómicos en que es posible la percepción de las abejas. Se propone que los distintos tipos de exposición de la luz estudiados son una parte importante en la experiencia, interacción, comunicación y alimentación de estos insectos.

## Introducción

La información que los seres vivos tienen del mundo externo no es perfecta. La imagen óptica de un objeto no llega al cerebro y es percibida sin ningún intermediario sino que está determinada por las condiciones físicas presentes en un gran campo óptico que va desde la fuente hasta el cuerpo que percibe. Las teorías tradicionales de la percepción más aceptadas tratan de una cadena causal de reacciones físicas.

- La luz, puede ser concebida como un fotón o un campo electromagnético continuo y ondulatorio emitida desde una fuente material (el sol) en cierta dirección.
- La luz se propaga en el espacio y los materiales de acuerdo con leyes y circunstancias físicas.
- La luz llega al ojo, se impregna en la retina y provoca que las células de la retina generen un potencial eléctrico.
- Este impulso eléctrico, junto a otras corrientes eléctricas e impactos de presión en el ojo, se propaga por el nervio óptico y el sistema visual nervioso.
- Se instauran complejos y desconocidos estados cerebro/mente y ocurre la experiencia visual. (Abraham, 2013.)

La mayor parte de la luz natural, del sol, no está polarizada. La dirección transversal del campo eléctrico en la luz natural varía rápidamente (y casi al azar). Dichas variaciones implican la superposición de múltiples frecuencias en oposición a la frecuencia única asumida en la formulación del cálculo de Jones. La característica principal de la luz no polarizada es que no se puede extinguir con un solo polarizador.

Las cualidades del color compuesto forman una variedad de infinitas dimensiones desde el punto de vista físico. Sin embargo, Las cualidades de color simples forman una variedad unidimensional. Para la descripción completa de un color compuesto se requiere la indicación con qué intensidad se representa cada una de las infinitamente posibles longitudes de onda λ; de modo que involucra infinitas variables independientes Jλ. Así, las ondas planas de dependencia temporal más generales pueden verse como superposiciones de componentes individuales de una sola frecuencia (a través de la transformada inversa de Fourier), y por lo tanto podemos considerar las ondas monocromáticas como bloques de construcción para ondas planas más complicadas.

Luego de que la luz ha sido transmitida a la percepción visual, la cadena causal continua. Mediante estudios de comportamiento animal se determinó que las abejas son susceptibles a la luz ultravioleta y al fenómeno de la recurrencia (Weyl, 1934), ya que, al privar a las abejas de este espectro de luz, las abejas pierden interés en alimentarse. Así mismo, al cambiar o privar la luz polarizada, las abejas pierden el sentido de ubicación. Posteriores estudios demostraron que los aparatos visuales de las abejas tienen las estructuras necesarias para percibir estos aspectos de la luz.

Este trabajo de física está enfocado en los primeros pasos de la cadena causal, principalmente en las propiedades como la polarización de la luz presente en el entorno de las abejas que podría ser relevante en su percepción visual e interacción con el entorno.

## Desarrollo y Marco Teórico
## La fuente de Luz

La luz solar blanca es la mayor fuente natural de luz. En esta sección, investigaremos algunas de las propiedades únicas de las ondas de luz provenientes del sol e introduciremos aspectos fundamentales como la longitud de onda y la polarización.

La luz no polarizada puede polarizarse parcialmente en el cielo dependiendo de la posición del sol o cuando, se refleja desde una superficie con incidencia oblicua, como en la superficie de un cuerpo de agua, ya que sus componentes pueden reflejarse con diferentes intensidades. (Peatross, 2015)

Para realizar un seguimiento de la polarización parcial (y la atenuación) de un haz de luz a medida que la luz progresa a través de un sistema óptico. Podemos considerar cualquier haz de luz como una suma de intensidad de luz completamente no polarizada y luz perfectamente polarizada

$$I = I_{pol} + I_{np}$$

y se define al grado de polarización como: la fracción de la intensidad que se encuentra en un estado de polarización definido

$$\xi_{pol} \equiv \frac{I_{pol}}{I_{pol} + I_{np}}$$

El grado de polarización adquiere valores entre cero y uno. Es 0 cuando la luz es completamente no polarizada y es 1 cuando el rayo es completamente polarizado.

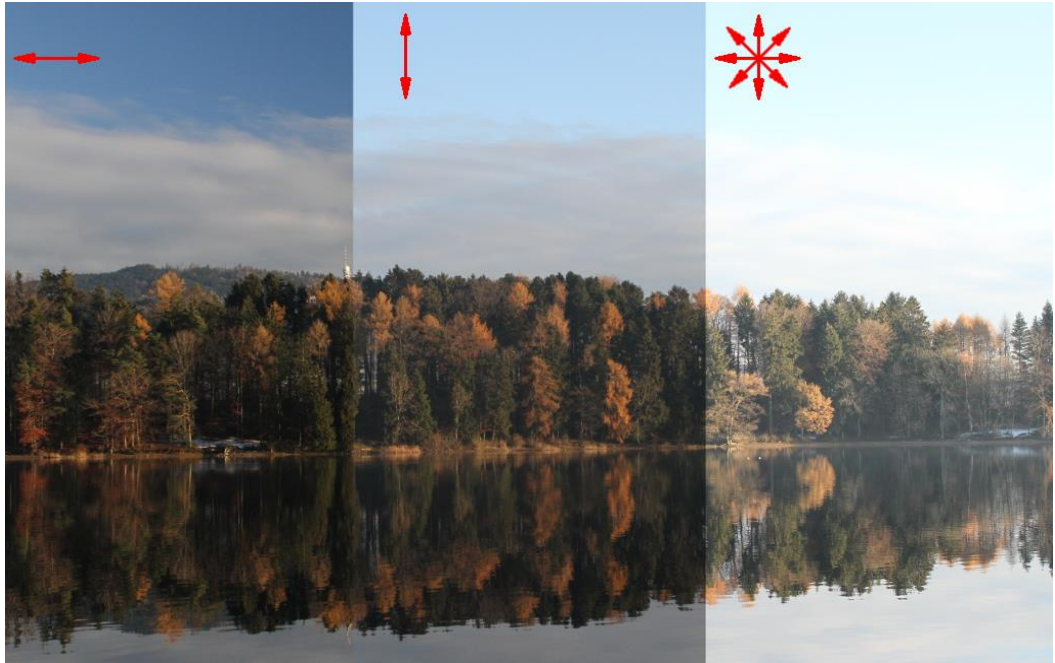

*Figura 1: Luz parcialmente polarizada, con un polarizador horizontal (que bloquea parte de la luz azul polarizada verticalmente), vertical y ninguno*

Luego, se asume que ambos tipos de luz se propagan en la misma dirección ya que los campos bidimensionales tienen una estructura simple en comparación con los campos tridimensionales, y esta estructura a menudo permite una descomposición en estructuras de campo aún más simples.

Una propiedad importante de la onda plana uniforme es la distancia entre frentes de onda adyacentes que producen el mismo valor de la función coseno. Si permite a $r_1$ y $r_2$ ser puntos en frentes de onda adyacentes.

$$\lambda = \mathbf{k} \cdot (\mathbf{r_2} - \mathbf{r_1}) = r_{02} - r_{01}$$

$\lambda$ se define como la longitud de onda y solo tienen significado para los campos de luz monocromáticos. Todos los efectos físicos esta luz están completamente determinado por la longitud de onda junto con la intensidad (Weyl, 1934). Las ondas monocromáticas pueden transportar energía y son un caso especial en el que el espectro de la señal de onda plana consiste en un único componente de frecuencia.

El hecho de que los seres vivos de la Tierra ven las cosas con precisión a longitudes de onda de alrededor de 600 nanómetros está indudablemente relacionada con el hecho de que, la luz solar (cuando se descompone espectralmente) tiene su máxima intensidad en ese valor. (Weyl, 1934). Los humanos generalmente vemos un espectro de 750 a 400 nm, mientras que se estima que las abejas pueden ver un rango de 600 a 300 nm que incluye la luz ultravioleta.

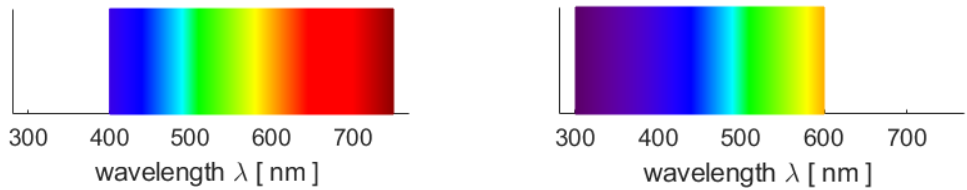

*Figura 2: Espectro de luz visible humano vs el espectro de la abeja.*

## La Polarización de una onda plana

La polarización de onda plana describe la evolución temporal de la dirección del vector del campo eléctrico, que depende de la manera en el que se genera la onda. Las ondas parcialmente polarizadas se generan en las fuentes de radio cósmicas y en muchas estructuras naturales como en los reflejos de la superficies.

Como ya se mencionó, las ondas parcialmente polarizadas contienen componentes completamente polarizadas y no polarizadas. Un vector de Stokes, que caracteriza una onda plana parcialmente polarizada, se escribe

$$[S_0, S_1, S_2, S_3]$$

$$S_0 = \frac{I_{pol} + I_{np}}{I_{in}}; \quad S_1 = \frac{2I_{hor}}{I_{in}} - S_0; \quad S_2 = \frac{2I_{45°}}{I_{in}} - S_0; \quad S_3 = \frac{2\,I_{r-cir}}{I_{in}} - S_0$$

El parámetro $S_0$, compara la intensidad (o potencia) total del haz de luz con un punto de referencia o intensidad de entrada, medida antes de que el haz ingrese al sistema óptico. Representa la intensidad en el punto de investigación, donde se desea caracterizar el haz.
$S_1$, describe la cantidad de luz polarizada horizontal o verticalmente
$S_2$ describe la cantidad de luz polarizada linealmente en las diagonales.
$S_3$ describe la cantidad de polarización elíptica y circular en el rayo de luz.

El matemático Henry Poincaré se dio cuenta de que los parámetros de Stokes ($S_1$, $S_2$, $S_3$) describen un punto en una esfera de radio $S_0$, y que esta esfera de Poincaré es útil para visualizar los diversos estados de polarización. Cada estado corresponde únicamente a un punto en la esfera. Por lo tanto, podemos mapear los estados de polarización directamente en la esfera: las polarizaciones izquierda y derecha aparecen en los hemisferios superior e inferior, respectivamente; la polarización circular aparece en los polos; la polarización lineal aparece en el ecuador. Los ángulos α y δ también tienen interpretaciones geométricas.

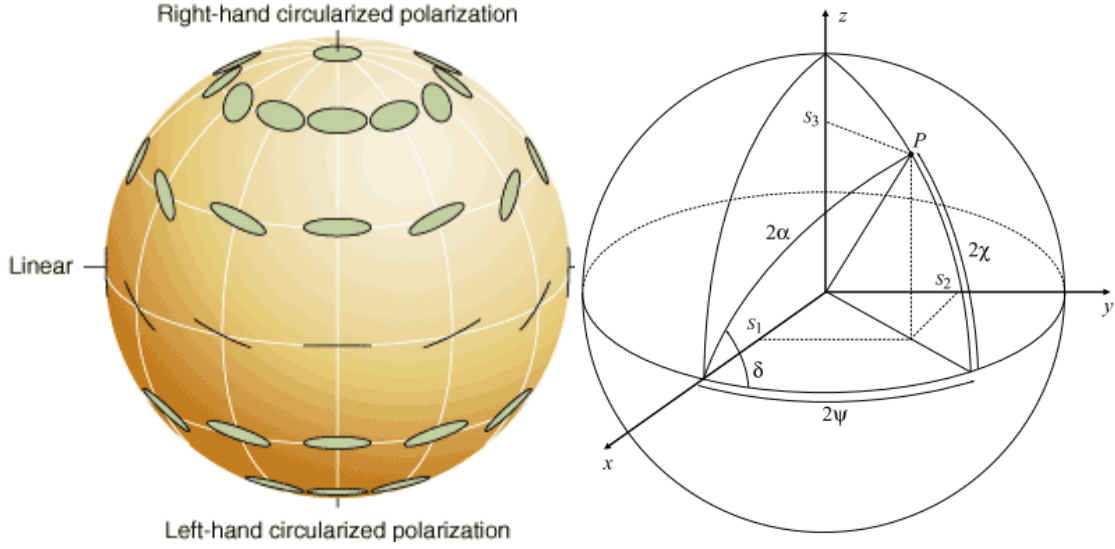

*Figura 3: Representación gráfica de la polarización de una onda plana en la esfera de Poincaré*

Si tomamos en cuenta la evolución temporal del campo eléctrico de la onda plana se tiene

$$\mathbf{E}(\mathbf{r}, t) = \mathrm{Re}\left\{\check{\mathbf{E}} e^{j\omega t}\right\} = \hat{\mathbf{x}} E_{x0} \cos(\omega t - kz + \phi_x) + \hat{\mathbf{y}} E_{y0} \cos(\omega t - kz + \phi_y)$$

Y se puede escribir una forma completamente general de el vector de Stokes como

$$s_0 = \frac{1}{2\eta}\left[E_{x0}^2 + E_{y0}^2\right],$$

$$s_1 = \frac{1}{2\eta}\left[E_{x0}^2 - E_{y0}^2\right] = s_0 \cos(2\chi)\cos(2\psi),$$

$$s_2 = \frac{1}{\eta} E_{x0} E_{y0} \cos\delta = s_0 \cos(2\chi)\sin(2\psi),$$

$$s_3 = \frac{1}{\eta} E_{x0} E_{y0} \sin\delta = s_0 \sin(2\chi).$$

Es un caso interesante cuando χ=pi/4 por lo que δ=pi/2 que resulta una polarización circular

$$\mathbf{E} = E_{x0}\left[\hat{\mathbf{x}} \cos(\omega t - kz) \mp \hat{\mathbf{y}} \sin(\omega t - kz)\right]$$

Muchos insectos reflejan luz polarizada circularmente (Arwin, 2012)

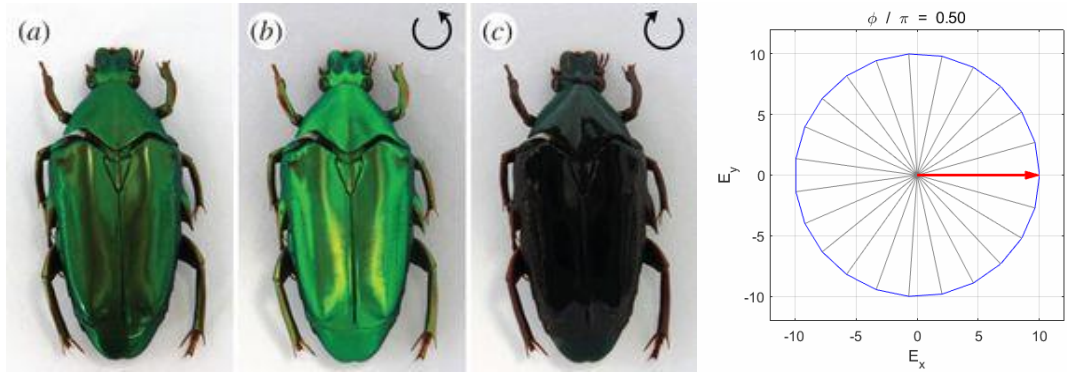

*Figura 4: Insecto visto normalmente, con un polarizador circular antihorario y horario.*

Otro caso a considerar es cuando χ=0 por lo que δ=0 y resulta una polarización lineal en donde el vector de campo eléctrico traza una línea recta.

$$\mathbf{E} = \left(\hat{\mathbf{x}} E_{x0} + \hat{\mathbf{y}} E_{y0}\right) \cos(\omega t - kz + \phi_x).$$

Si E$_{xo}$=0 es una polarización linear vertical y si E$_{yo}$=0 es horizontal. Las abejas saben dónde están las fuentes de agua debido a que cuando los rayos solares reflejan sobre la superficie de un charco o un lago se reflejan una gran cantidad de ondas polarizadas horizontalmente.

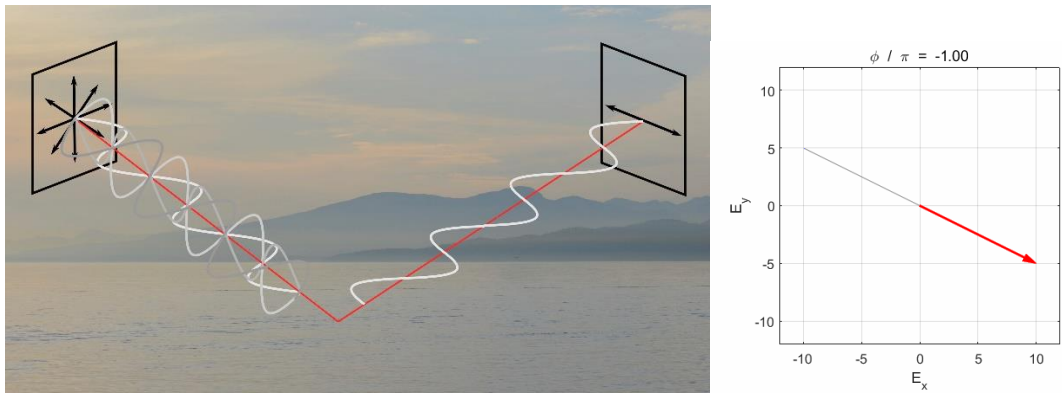

*Figura 5: Polarización lineal horizontal y polarización lineal con dos componentes*

Si los rayos solares inciden con el ángulo de Brewster θ$_B$ la componente de la polarización paralela al plano de incidencia se anula en el haz reflejado. Este ángulo de Brewster que hemos introducido depende de la permitividad eléctrica y permeabilidad magnética entre los medios. Es frecuente que las permeabilidades magnéticas de ambos medios no varíen y en este caso, el ángulo de Brewster se puede calcular a partir de los índices de refracción de ambos medios. Utilizando la ley de Snell se obtiene se demuestra que,

$$\tan(\theta_B) = \frac{n_2}{n_1}$$

La polarización por reflexión es máxima en el ángulo de Brewster, cuando la tangente del ángulo de incidencia es igual al índice de refracción de la sustancia. En el caso del agua el ángulo de Brewster es $\theta_B = tan^{-1} 1.33 = 53.06°$ La polarización es nula para la incidencia perpendicular.

Las hojas y pétalos de las plantas (Kelber, 1999) por sus superficies lisas producen reflejos polarizados. Además que muchas de ellas reflejan luz ultravioleta.

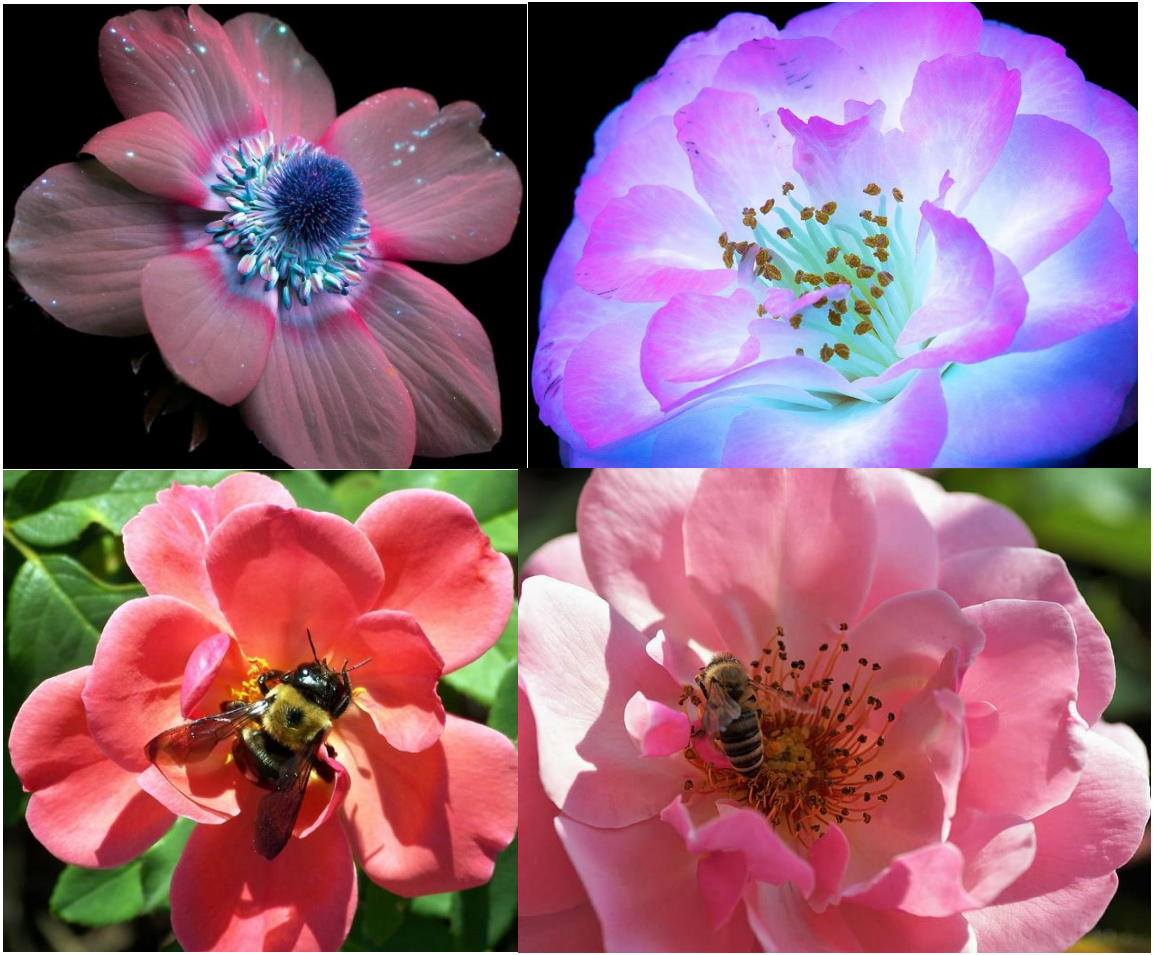

*Figura 7: Flores bajo luz ultravioleta (tomadas por Burrows C.) que simulan la visión de las abejas y fotos de abejas con iluminación normal*

El sistema visual de la abeja le permite distinguir los colores de la luz ultravioleta. Los estudios comportamentales confirman esto. La visión ultravioleta de las abejas les sirve para encontrar polen y néctar en ciertas flores son como luces guías en el néctar que estos insectos pueden ver aunque sean invisibles para los humanos.

Además de este fenómeno de reflexión de una onda plana en una interfaz de material plano, es posible estudiar la reflexión en cualquier número de capas de material, entre diferentes regiones materiales y en diferentes interfaces que producirán distintos tipos de polarización que seguramente las abejas reconocen en su percepción, sin embargo puede ser complejo y excede los propósitos de este trabajo.

Uno de los fenómenos más importantes que ocupa los conceptos ya estudiados repercute en la ubicación diaria de las abejas. La fuente más amplia de luz es el sol, entra en la atmósfera superior se dispersa hacia un observador terrestre por centros de dispersión (átomos, moléculas, aerosoles, fluctuaciones de la densidad del aire) más pequeños que la longitud de onda de la luz. (Foster, 2017) La luz no polarizada del sol incide e interacciona con las partículas del aire. En las moléculas se crea un momento de dipolo, este dipolo no es capaz de emitir en su misma dirección. Por ejemplo, Supongamos que la luz solar está en el plano x-y y se propaga en el eje z, la componente vertical y genera un momento dipolar vertical que no emite en es dirección por lo que resulta polarizada horizontalmente en x.

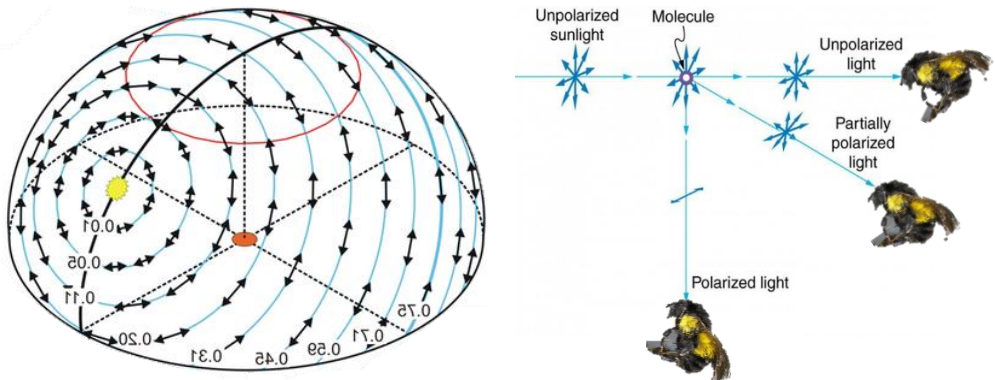

*Figura 6: Polarización atmosférica*

Los distintos tipos de polarización resultantes de esta interacción son usados por las abejas para determinar la posición del sol y tener una orientación confiable incluso cuando el sol no es directamente visible (Horváth, 2014) y cuando la luna es la fuente primaria de luz durante las noches.

## La Percepción de las ondas de luz

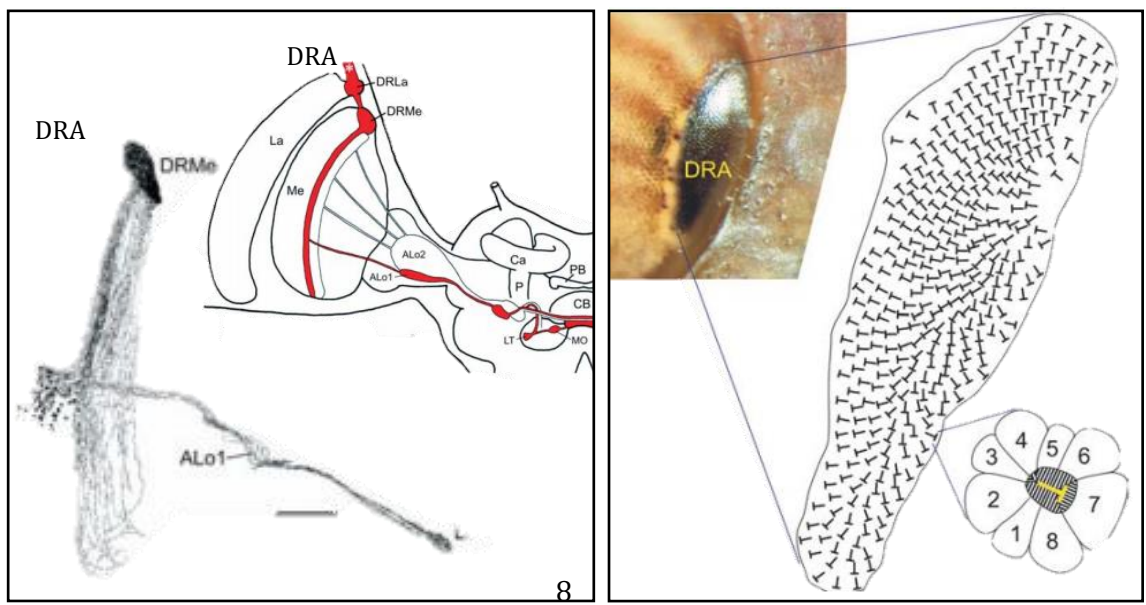

*Figura 8: Fotorreceptor del área del borde dorsal del ojo (DRA) sensible a la polarización proyectando un impulso en la vía óptica. Figura 9: Área dorsal del ojo de la abeja sensitiva a la polarización de la luz y a la luz ultravioleta, se muestra un arreglo de omatidias formadas por células fotorreceptoras*

El ojo de la abeja es capaz de medir la intensidad de la luz al igual que un analizador. La luz es el principal estímulo de los fotorreceptores pero además hay otros estímulos como la presión. Cada omatidio tiene 8 células fotorreceptoras, cuatro células visuales responden óptimamente a la luz amarillo-verdosa (544 nm); dos responden a la luz azul (444 nm) y las dos restantes responden a la luz ultravioleta (344 nm). Las microvellosidades se ubican perpendicularmente a cada una de estas células. La sensibilidad a la polarización en los fotorreceptores de invertebrados se basa en las propiedades de absorción de luz polarizada de las microvellosidades (Goldsmith y Wehner, 1977). Las moléculas de pigmento visual están alineadas dentro de la membrana microvillar de tal manera que la luz polarizada linealmente se absorbe al máximo cuando el plano de oscilación (del vector E) es paralelo al eje largo de las microvellosidades.

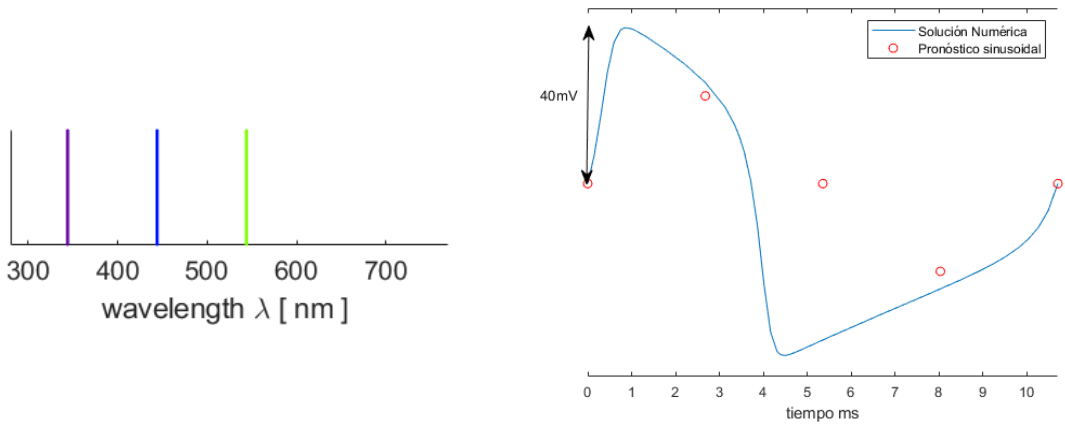

*Figura 10: Longitud de onda que de los receptores visuales. Figura 11: Simulación de la acción potencial de una célula fotorreceptora del ojo de la abeja ante un estímulo lumínico.*

El estímulo lumínico provoca una disminución de la resistencia de toda o parte de membrana celular de la retina, dando lugar al potencial de acción. El potencial de acción provoca la salida de la corriente a través de la célula, esta corriente de despolarización inicia los potenciales de pico del proceso proximal. (Kén-ichi, 1962) El impulso eléctrico, se propaga por el nervio óptico y el sistema visual nervioso.

Finalmente instauran complejos y desconocidos estados cerebro/mente donde la abeja adquiere la información lumínica. Esta información le permite a la abeja moverse por el mundo y dar respuestas como danzas que sirven para orientar al resto de la colmena sobre los lugares donde pueden encontrar agua o alimento.

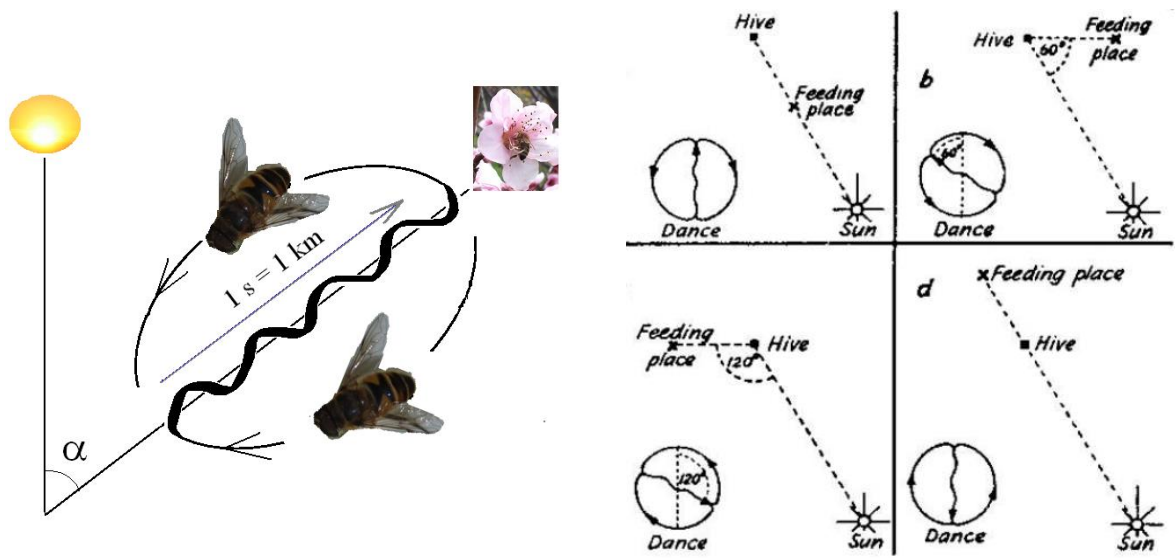

*Figura 12: Danza de las abejas que toma en cuenta la ubicación de sol y da información a la colmena sobre cierta ubicación.*

## Conclusiones y observaciones

En este proyecto se estudió la cadena causal de la percepción de las ondas de luz en las abejas, esta no va directamente de la fuente al individuo si no que su transición depende de varios estados físicos. Se utilizaron conceptos de campos electromagnéticos, conservación de la energía, intensidades, longitudes de onda, espectros de luz, polarización y señales eléctricas. Mediante la descripción detallada de la polarización de la luz se dio explicación a varios fenómenos naturales con los que están en contacto las abejas. Existen una gran cantidad de fuentes de polarización en el entorno natural que vale la pena analizar ya que incluso pueden ser una parte importante de la comunicación e interacción animal. Así mismo, experimentalmente se deben considerar más experimentos en los que los animales están expuestos a la luz polarizada y a otros espectros de luz.

## Referencias